\documentclass [12pt]{article}
\usepackage {graphicx}
\usepackage {epsfig}
\begin{document}

\title{\Large Frequency dependence of the intergranular
magnetic flux penetration in ceramic  $YBa_2Cu_3O_x$ superconductor}

\author{A.A.Sahakyan, S.K.Nikoghosyan, G.N.Yeritsyan\\
\small Solid States Radiation Physics Laboratory, Yerevan Physics Institute, Armenia}
\maketitle

\bigskip
{ \small By means of small magnetic field ac susceptibility measurement
at 10 kHz we found that the real and imaginary parts of ceramic
$YBa_2Cu_3O_ x$ susceptibility in presence of the external low
frequency field close to 0.1 Hz exhibit frequency dependence. The wide
maximum of hysteresis losses and  the exponential
dependence of the effectiveness of flux penetration
with increasing of external field frequency were obtained. We observed
a nonlinear dynamic magnetic response in presence dc field and we suggest that
this behavior is due to the dynamic and static Josephson vortex-vortex
interaction.}

 \section{Introduction}

 In an early stage of studies of the ac magnetic field
response in granular high temperature superconductors (HTSC) many
researchers have reported that the temperature $T^J_ m$, indicating the
maximum of hysteresis losses of complex magnetic susceptibility $\chi=
\chi' -i\chi''$  shifts to the lower values at the applied dc magnetic
field on the sample [1-3]. Lobotka and Gomory have  reported that the
larger dc field values than ac field amplitudes are needed to shift the
$\chi''$  peak temperature $T^J_m$ by the same amount [1]. According to
Clem [4] and Gershkenbein, Vinokur, and Fehrenbache [5] the
 $T^J_m$ in $\chi''$ corresponds to the temperature at which
ac magnetic field with amplitude $h_{ac}$ fully penetrates the sample
volume. From this point of view one can expect a frequency dependence
of  magnetic flux penetration effectiveness in intergranular medium of
ceramic superconductor. In [6] some increase of  $T^J_m$ ($\Delta T^J_m
\cong 2$ K)  with frequency in the frequencies range of ac magnetic
field from 300 Hz to 80 kHz was found and authors explained it by a
spin-glass-like behavior of superconducting granular HTCS sample. Other
authors [7-10] pointed a weak dependence or full absence [11-13] of
frequency dependences of complex magnetic susceptibility in ceramic
HTSC materials.

In this paper we present the results of experimental investigations of
the frequency dependences of magnetic field penetrations in ceramic
$YBa_2Cu_3O_x$ superconductors at non full magnetic flux penetration
regime of the sample volume at temperature T=80 K for both external
sinusoidal magnetic field in frequency range from $f_{ex}=0.01$Hz to 90
Hz and amplitude $H_{ac}\le 40$ Oe and static magnetic field
$H_{dc}\le40$ Oe. In our experiments we used the small magnetic field
(amplitude 2 mOe) ac susceptibility measurements technique at frequency
10 kHz.

The real part of ac susceptibility has been taken as a deviation
 $\Delta \chi' =(\chi'_0 -\chi')/\chi'_0 $, where $\chi' $
 is the current value of the  real
part  and $\chi'_0$ is the value of $\chi'$ at which a full screening
state of the sample occurs. The magnitude of $\Delta \chi' $ is
proportional to the fraction of the sample volume where the Josephson
vortex penetrated. $\Delta \chi' =0$ corresponds to full screening
state of the sample and our measurement $\Delta \chi'=0.28$ corresponds
to the sample state at which Josephson vortexes reaches the center of
the sample (see. Fig.1(b) in the text).

\section{ Experimental}

 Ceramic  $YBa_2Cu_3O_x$  samples with cylindrical shape
of 3mm in diameter and 8mm long were prepared by standard solid-state
reaction method. The real $\chi' $ and imaginary $\chi''$ parts of
complex magnetic susceptibility of HTSC samples were measured by small
magnetic field ac susceptibility measurements technique at frequency 10
kHz and amplitude 2 mOe in presence both static field or low frequency
sinusoidal magnetic field. The ac magnetic field is considered small if
its contribution to the measuring values of the susceptibility
components $\chi' $  and $\chi''$ are  negligible. For each sample and
measuring regime exist experimentally determined range of small
measuring field amplitude $h_{ac}$ at which both real and imaginary
components of ac susceptibility are independent of $h_{ac}$. On the
other hand the reduction of the ac field amplitude is limited by the
equipment sensitivity. Taking into account these two requirements the
optimum ac field amplitude $h_{ac}$ is the high end of the small
magnetic field range. For our investigations $h_{ac}=2$ mOe was
obtained (see. Fig.2 in the text).

Complex magnetic susceptibility was measured using a homemade ac
inductance bridge working at frequency $f_ m=10$ kHz with amplitude
$h_{ac}$ in range 0.5 mOe to 10 Oe. The real $\chi' $ and imaginary
$\chi''$ parts of the ac complex susceptibility are determined from the
data of measuring coil $L_{ m}$ parameters deviation caused by a
superconducting sample  which were surrounded by this coil. The $L_m$
is connected in the arm of the ac inductance bridge. Another external
larger coil $L_{ex}$ was used for creation of static $H_{dc}$ or
sinusoidal $H_{ex}(t)$ magnetic field in frequency rang from 0.01 Hz to
90 Hz and amplitude up to 40 Oe, where t is the time. The axes of both
coils were orientated parallel to the cylindrical sample axis. The
phase angle adjustment of two Lock-in detectors for both real and
imaginary signal components of ac bridge was carried out connecting and
disconnecting a calibrating inductance and resistance in series to the
measuring coil $L_m$.

In our experiments the influence of the Earth magnetic field on
measured values of the real $\chi' $ and imaginary $\chi''$ parts
 was negligible and all measurements in ambient Earth
magnetic field were carried out. Temperature of the sample was
monitored with help of a copper wire resistor with relative accuracy of
about 0.1 K. The measuring coil $L_m$ was put in the liquid nitrogen
and the measurements were carried out in heating regime of the sample
at temperature rate about 1 K/min. Our experimental data have appeared
reproducible within the sensitivity of our measurements for all
investigated samples and their measuring regimes.

\section{Experimental results and discussions}

In Fig.1 (a) and (b) we show the temperature dependences of the
imaginary $\chi''$ and real $\chi'$ components of the ac complex
susceptibility $\chi(T)=\chi'(T)-i\chi''(T)$ of the ceramic sample
$YBa_2Cu_3O_x$  measured for different ac field amplitudes and
superimposed dc field regimes. The curves 1, 2 and 3 in this figure
corresponds to the measuring ac magnetic field amplitude $h_{ac}=2$
mOe, 0.9 Oe and 5,3 Oe respectively, and they have not noticeable
features and are similar to the results previously reported by many
researchers. However there are many interesting peculiarities at
comparison on the curves 2(a), 2(b) and 4(a), 4(b) in a Fig.1, where
curve 4(a) and 4(b) are obtained by measuring ac susceptibility at
amplitude $h_{ac}=2$ mOe in the presence of dc field $H_{dc}=40$ Oe. In
the presence of dc field the curves 1(a) and 1(b) in the Fig.1 are
transformed and the temperatures $T^J_m$ of hysteresis losses peak
$\chi'' _{\rm max}$ are  shifted to lower temperatures and the slope of
a curve $\chi'(T)$ decreases strongly.

In order to compare the ac and static $H_{dc}$ magnetic fields flux
penetration effectiveness into the sample volume, from the $\chi'(T)$
and $\chi''(T)$ curve families the curves for which the hysteresis
losses peak temperature $T^J_m$ have the same value (curves 2(a) and
4(a) in Fig.1) were chosen, i.e. in both cases the magnetic flux front
reaches the sample center at temperature  $T^J_m =87,2$ K. In case of
the presence of a dc field $H_{dc}=40$ Oe the contribution
 from a measuring magnetic field $h_{m}(t)$ with frequency 10 kHz and
 amplitude 2 mOe to the
value of $\chi (T)$  is negligibly small, because in the amplitude
range from 0,5 mOe up to 2 mOe the $\chi'(T)$ and $\chi''(T)$
 at the temperature interval from 80 K to 92K do not vary
noticeably. It means that in the latter case the magnetic response is
caused by a static magnetic field.

The obtained experimental results show, that the full penetration
regime of the magnetic flux in the sample volume at temperature T=87,2
K in both cases take place at the ratio of dc field to the effective
value of the ac field equal to $\sqrt{2}H_{dc}/h_{ac}\cong 60 $.
Therefore we can suggest that the effectiveness of the magnetic flux
penetration in sample volume for static and ac fields strongly differ.
We note that this ratio decreases with increase of $T^J_m$ and at
T=88,3 K is equal 50. The decrease of $T^J_m$ in $\chi''$ at a
superimposed dc field were reported in [1-3], however the contributions
to the measured values of $\chi' $ and $\chi''$ of static field and ac
field were not separated and  at that case it is very difficult to
judge on penetrating ability of dc field and ac field in the sample
volume.

 Thus, our experimental data show that the Josephson
vortexes in  $YBa_2Cu_3O_x$  ceramic samples created by static and ac
magnetic fields have a different character and we named them static and
dynamic Josephson vortex respectively. As can be seen by comparing
curve 2(a) and curve 4(a) in Fig.1, the penetration effectiveness of
dynamic vortexes in intergranular medium of HTSC ceramic
superconducting sample is much more than that for a static vortex and
therefore the density of dynamic Josephson vortexes is less than static
Josephson vortexes density.

 Another feature of the curve 4(a) in comparison with
curve 2(a) in Fig.1 is that from temperature T=83 K and higher the
hysteresis losses $\chi''$ at the presence of static field $H_{dc}=40$
Oe are noticeably lower than those for $\chi''$ measured by ac field at
frequency 10 kHz with amplitude  $h_{ac}=0,9$ Oe. We suggest that such
behavior of the magnetic response is due to vortex - vortex
interaction. In other words hysteresis losses at interaction of dynamic
vortexes of the measuring small field with the static vortexes are less
than hysteresis losses occurring at interaction of dynamic vortexes
with each other at measuring field amplitude $h_{ac}=0,9$ Oe.

 In the Fig.1(b) the temperature dependences of a
real part of complex magnetic susceptibility $\chi'(T)$
 measured at different field regimes are shown. As
can be seen in Fig.1 the behaviors of $\chi' (T)$ in curve 2(b) and
curve 4(b) in the neighborhood of temperature $T_m^J$ differ slightly,
whereas at the same temperature range the differences of $\chi''(T)$
are clearly visible. From the Fig. 1 we can see that the hysteresis
losses peak $\chi''_{\rm max}$ for all measuring regimes occurs at
value of $\chi' \cong -0.72$ and this corresponds to $\Delta\chi '
\cong 0,28$.

 The difference in behavior of the magnetic response in
HTCS samples in the presence of dc and ac magnetic fields is visible
not only at temperature rang of occurrence of histeresis losses peak,
but it is clearly seen in the behavior of $\chi'(f_{\rm ex})$ and
$\chi''(f_{\rm ex})$ for low frequency external magnetic field $H_{\rm
ex}(t)$. However, before presenting these data, it is necessary to show
that the values of $\chi'(f_{\rm ex})$ and $\chi''(f_{\rm ex})$ are
caused by action of $H_{ex} (t)$ only, because the ac susceptibility
measurements will be carried out at frequency $f_{\rm m}=10$ kHz. To do
this it is necessary to determine experimentally the range of small
measuring magnetic field amplitude and further, to find its optimal
magnitude.

 With this purpose both the real $\Delta\chi ' $ and
imaginary $\chi''$ components of complex susceptibility were measured
versus ac field amplitude $h_{\rm ac}$ in range from 0,5 mOe up to 10
mOe with frequency $f_{\rm m}=10$ kHz for different values of external
magnetic $H_{\rm ex}(t)$ field at the temperature T=80 K. These data
are presented in Fig.2 (a) and (b), where it is shown that in presence
of dc field, starting from the amplitude value $h_{\rm ac}=3$ mOe and
above, there is a noticeable increase of the imaginary $\chi''$ and
real $\Delta\chi ' $ components (curves 1,2,3 in a Fig.2 (a) and (b)).
It is also seen in this figure, that for fixed value of amplitude
$h_{\rm ac}$ the increases of the magnitudes of these components depend
on static field value and that the higher the magnitude of $H_{\rm
dc}$, the more is the magnitude of contribution of the measuring field
to these components. In the case $h_{\rm ac}$=10 mOe both components
rise almost by factor two whereas at the absence of the dc field the
magnitude of imaginary $\chi''$ and real $\Delta \chi' $ parts are
equal to zero and remain independent on $h_{\rm ac}$ in the range from
0,5 mOe up to 10 mOe at the temperature of sample T=80 K.

 This is the most important feature from which we
can see that even in the case when the sample state is essentially far
from the magnetic flux fully penetration regime ($\Delta \chi '<0.04$)
the magnitude of imaginary $\chi''$ and real $\Delta \chi' $ components
of magnetic susceptibility are not a superposition of the contributions
of each of these fields, and have more complex character. We suggest
that the appearance of such nonlinear magnetic response in this case is
caused by the interaction of the dynamic vortexes of ac measuring field
with the static vortexes of dc field  and this effect becomes
noticeable at $h_{ac} \ge 3 $ mOe. In other measuring regime, when the
external field has the sinusoidal shape, the dependence of
$\Delta\chi'$ and $\chi '' $ on $h_{\rm ac}$ practically is not
noticeable in the range of  $h_{\rm ac}$ from 0,5 mOe up to 10 mOe (see
curves 4, 5, 6 in a Fig.2 (a) and (b)). In the last case we suppose
that the interaction of two dynamic incoherent Josephson vortexes
created by the measuring ac field with frequency $f_{\rm m}$=10 kHz and
amplitude up to 10 mOe with an external field $H_{\rm ex}(t)$ at
frequency  $f_{\rm ex}\le 90$ Hz does not give noticeable contribution
to the values $\Delta\chi'$ and $\chi''$

 Thus the experimental data in a Fig.2 (a) and (b)
indicated that: a) the amplitude  $h_{\rm ac}$=2 mOe of the low
magnetic field at frequency 10 kHz is optimal for measuring values of
the real $\Delta\chi'$ and imaginary $\chi''$ components and therefore
it is possible to state that the magnetic response in ceramic samples
in our investigations was caused by external field \ $H_{\rm ex}(t)$
only; b) the interaction of the dynamic vortexes with the static
vortexes, dynamic vortex-vortex and dynamic incoherence vortexes have
different behavior. From obtained experimental data we can suggest that
the dynamic magnetic response of ceramic $YBa_2Cu_3O_x$ superconductor
in none fully magnetic flux penetration regime in intergranular
environment is substantially determined by three types of vortex-vortex
interactions. Note that it is necessary to take into account as well
the superconducting properties of the HTCS sample (pinning centers),
because during the penetration of dynamic Josephson vortexes into of
the sample bulk a noticeable role can be attributed to the interaction
of dynamic vortexes with static vortexes, frizzed in the sample volume
at penetration of a sinusoidal magnetic field after the first quarter
of the period. More detailed experimental data about vortex-vortex
interaction and the penetration dynamics of Josephson vortexes in
intergranular environment of HTCS sample will be introduced elsewhere.

 Fig.3 (a) and (b) show a frequency dependence of
imaginary  $\chi''$ and real $\Delta\chi'$  components of complex
susceptibility for some values of external magnetic field $H_{\rm
ex}(t)$ at the temperature T=80 K, measured by ac  field with the
amplitude 2 mOe and frequency 10 kHz.

 As can be seen the magnitude of hysteresis losses
 $\chi''(f_{\rm ex})$ at the presence of a static field and
a sinusoidal field up to frequency about $f_{\rm ex}=0.1$ Hz does not
exhibit a frequency dependence (Fig. 3 (a)). Further $\chi''$ starts to
rise and exhibits a wide maximum versus
 external magnetic field $H_{\rm ex}(t)$ frequency and
  the higher amplitude of a magnetic field, the
lower the frequency of occurrence of the  $\chi''$ peak.  For example,
the value of  $H_{\rm ac}$ = 40 Oe corresponds to the
 $\chi''$ peak frequency  $f_{{\rm ex}}^{\rm max}\cong 0.5$ Hz,
  and the value  $H_{\rm ac}=10$ Oe
corresponds to $f_{{\rm ex}}^{\rm max} \cong 5$ Hz.

Fig.3(b) shows that the magnitude of $\Delta\chi'$ up to frequency
about $f_{{\rm ex}}=0.1$ Hz does not depend on the frequency of
$H_{{\rm ex}}(t)$, but with the increase of $f_{{\rm ex}}$, for each
amplitude value of magnetic field $H_{\rm ac}$, there is a threshold
frequency, starting from which the $\Delta\chi'$ exhibits a frequency
dependence. The increase of $\Delta\chi'$ in this frequency range is
described by exponential law $\Delta\chi'\propto  f_{\rm ex}^{0.62}$. Thus
the increase
 of the frequency of the external magnetic field from $f_{{\rm ex}}\cong 0.1$
 Hz for fixed amplitude $H_{\rm ac}$ leads to the increases of the
effectiveness of the magnetic flux penetration  in the sample volume.

 Finally we note that qualitatively similar experimental
results were obtained for the bismuth based ceramic HTCS sample
($Bi_{1.7}Pb_{0.2}Sb_{0.1}Sr_{2}Ca_{2}Cu_{3}O_{x}$; $T_{{\rm c}}=108$
K) at the temperature T=80 K.

\section{Conclusions}

 We have shown that the real and imaginary parts of the
small magnetic field ac susceptibility at frequency 10 kHz and
amplitude 2 mOe in presence of low frequency external magnetic field at
non fully penetration regime in the ceramic  $YBa_2Cu_3O_x$ sample at
T=80 K close to $f_{{\rm ex}}=0.1$ Hz begin to exhibit frequency
dependence. With increasing of external magnetic field frequency the
wide maximum of hysteresis losses and  the exponential dependence
($\Delta\chi'\propto f_{\rm ex}^{0.62}$) of the  effectiveness of flux
penetration in the sample volume were obtained.

It is established that the penetration effectiveness of ac field at
frequency 10 kHz is much more than that for dc field and the magnetic
flux in both cases reaches the cylindrical sample center at temperature
T=87.2 K at the ratio of this magnetic fields $\sqrt{2}H_{{\rm
dc}}/h_{{\rm ac}} \cong 60$.

It is found that the dynamic magnetic response in ceramic
$YBa_2Cu_3O_x$  samples in presence of dc magnetic field is not a
superposition of the contributions of measuring ac field and dc field.
We suggest that in this case the nonlinear dynamic magnetic response
was caused by dynamic and static Josephson vortex-vortex interaction.\\

{\bf Acknowledgments}

 This work has been supported in part by
International Scientific and Technology Center under Grant \# A-102-2.

\pagebreak
 \vspace{5cm}
\begin{figure}
\centering
\includegraphics[width=14cm,height=10cm]{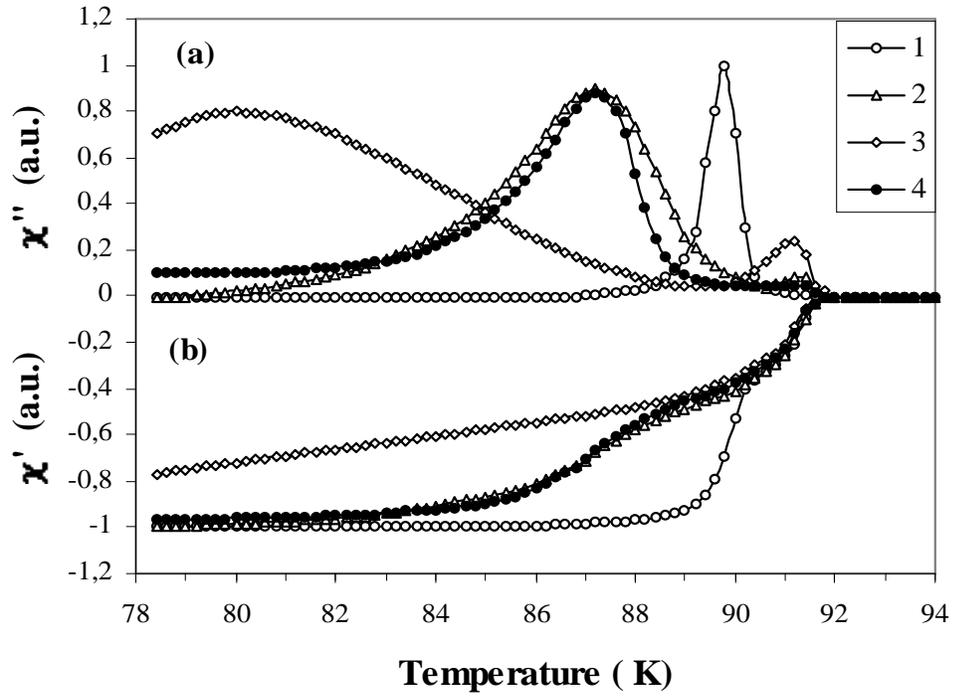}
\caption{Temperature dependences of the imaginary $\chi ''$ (a) and
real $\chi' $  (b) parts of ac susceptibility of the ceramic sample
$YBa_2 Cu_3 O_x$ at frequency 10 kHz for different measuring ac field
amplitude $h_{ac}$: 1 - 2 mOe, 2 - 0.9 Oe, 3 - 5.3 Oe, 4 - 2 mOe +
$H_{dc}$; $H_{dc}$=40 Oe.}
\end{figure}

\newpage

\begin{figure}
\centering
\includegraphics[width=12cm,height=10.2cm]{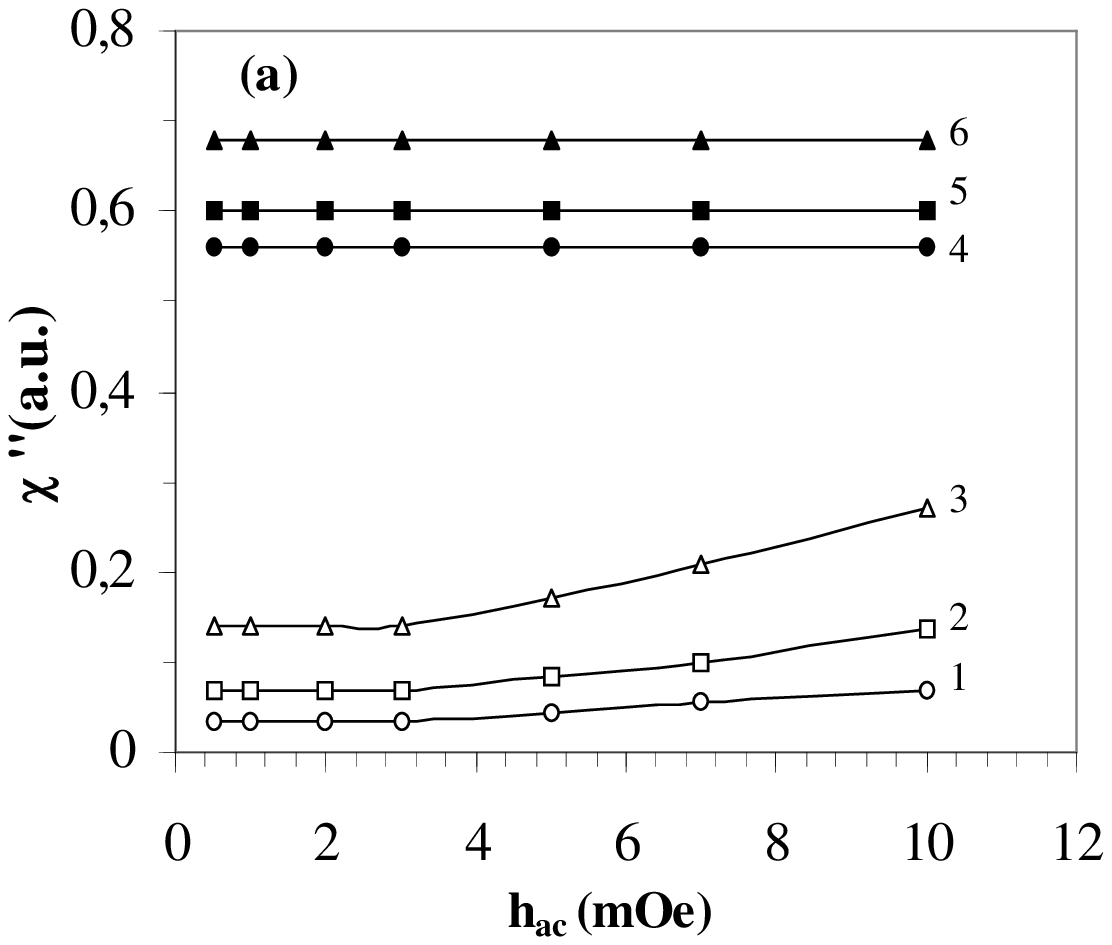}
\end{figure}
\begin{figure}
\centering
\includegraphics[width=13cm,height=9.6cm]{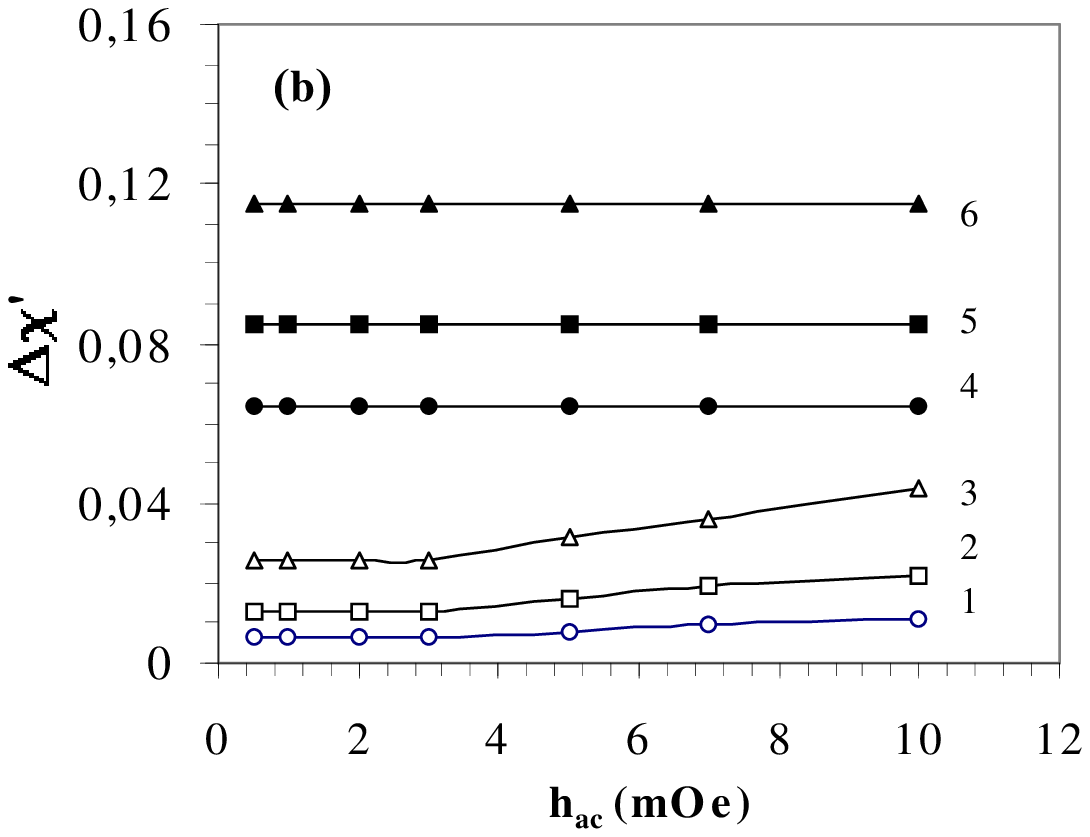}
\caption{ \small Dependence of imaginary $\chi''$ (a) and real $\Delta
\chi'=(\chi'_{o}-\chi')/\chi'_0$ (b) parts of $YBa_2Cu_3O_x$ sample
complex susceptibility versus measuring ac field amplitude $h_{ac}$ at
frequency $f_{m}=10$kHz at $T=80$ K in the presence of external static
field $H_{dc}$: 1 -- 7.1 Oe, 2 -- 14.2 Oe, 3 - 28,3 Oe; and external
sinusoidal field with amplitude $H_{ac}$ and frequency $f_{ex}$
respectively: 4 - 10 Oe, 5 Hz; 5 - 20 Oe, 2 Hz; 6 - 40 Oe, 1 Hz.}
\end{figure}

\newpage

\begin{figure}
\centering
\includegraphics[width=12cm,height=10cm]{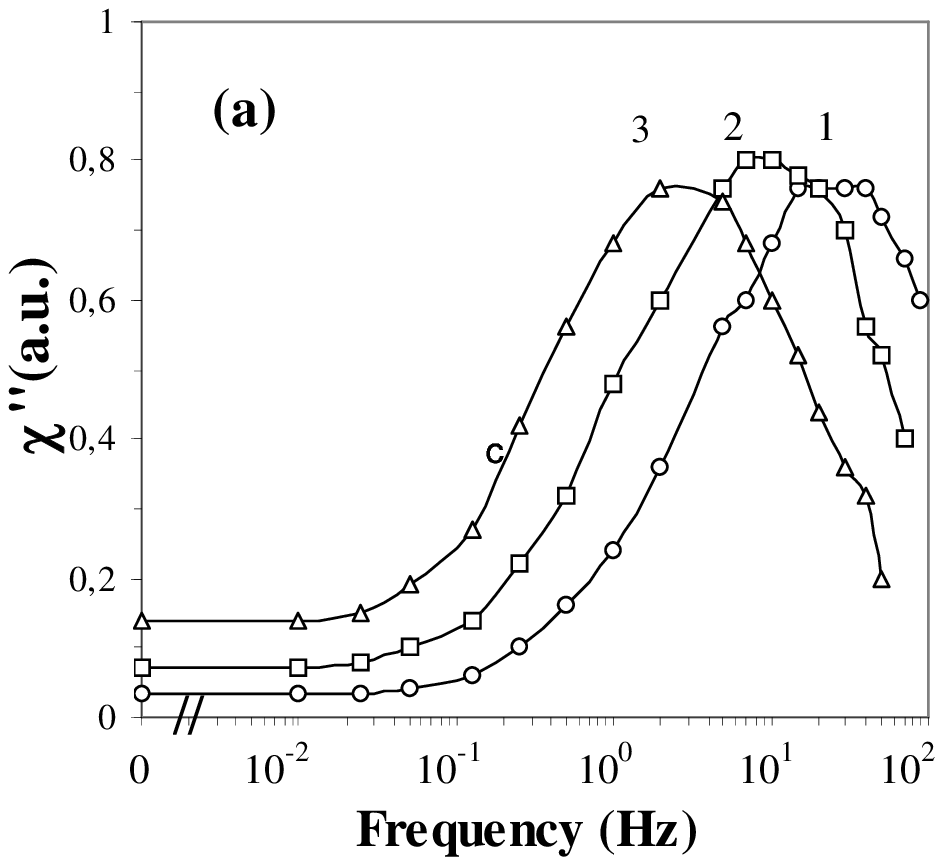}
\end{figure}
\vspace{-5cm}
\begin{figure}
\centering
\includegraphics[width=12cm,height=10cm]{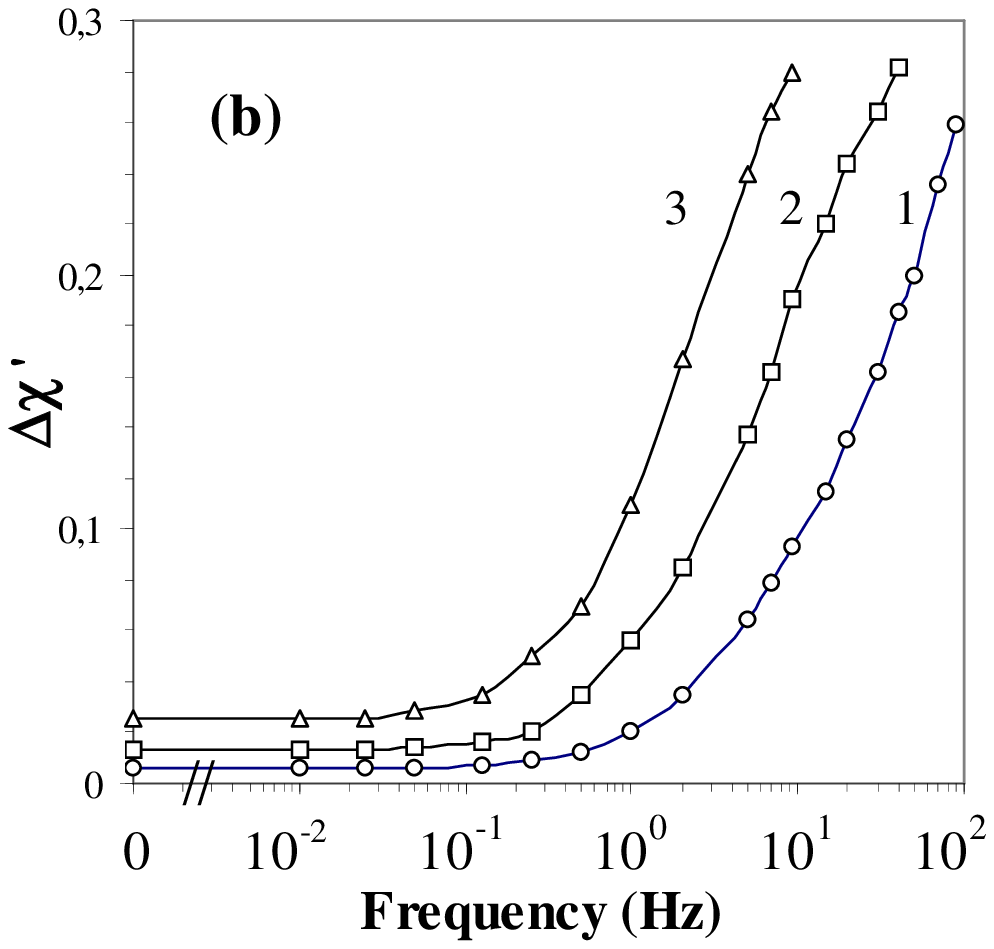}
\caption {\small Dependence of the imaginary $ \chi ''(a)$ and real
$\Delta\chi'=(\chi '_o-\chi')/\chi_0'$ (b) parts of ceramic sample
$YBa_2Cu_3O_x$ ac susceptibility versus external field frequency,
measured at frequency 10 kHz and amplitude 2 mOe at T=80 K for
different external field amplitudes $H_{ac}$: 1-10 Oe, 2 - 20 Oe, 3 -
40 Oe. The values of $\chi''$ and $\Delta\chi'$ at $f_{ ex}=0$
correspond to the dc field $H_{dc}= H_{ac}/\sqrt{2}$.}
\end{figure}

\end{document}